\documentclass{article}

\begin{document}

\title{Quarks and gravitation}
\author{G. Quznetsov \\
gunn@chelcom.ru, gunn@mail.ru}
\date{February 10, 2008}
\maketitle

\begin{abstract}
Turnings of the colors quarks mass space make curved the events time-space.
\end{abstract}

=============

In accordance with \cite{Four} the fermion moving equation has got the
following form:

\begin{equation}
\left( 
\begin{array}{c}
\beta ^{\left[ 0\right] }\left( \partial _0+\mathrm{i}\Theta _0+\mathrm{i}%
\Upsilon _0\gamma ^{\left[ 5\right] }-\mathrm{i}M_0\gamma ^{\left[ 0\right]
}-\mathrm{i}M_4\beta ^{\left[ 4\right] }\right) \\ 
+\beta ^{\left[ 1\right] }\left( \partial _1+\mathrm{i}\Theta _1+\mathrm{i}%
\Upsilon _1\gamma ^{\left[ 5\right] }+M_{\zeta ,0}\beta ^{\left[ 4\right]
}-M_{\zeta ,4}\gamma ^{\left[ 0\right] }\right) \\ 
+\beta ^{\left[ 2\right] }\left( \partial _2+\mathrm{i}\Theta _2+\mathrm{i}%
\Upsilon _2\gamma ^{\left[ 5\right] }+M_{\eta ,0}\beta ^{\left[ 4\right]
}+M_{\eta ,4}\gamma ^{\left[ 0\right] }\right) \\ 
+\beta ^{\left[ 3\right] }\left( \partial _3+\mathrm{i}\Theta _3+\mathrm{i}%
\Upsilon _3\gamma ^{\left[ 5\right] }-M_{\theta ,0}\beta ^{\left[ 4\right]
}-M_{\theta ,4}\gamma ^{\left[ 0\right] }\right)
\end{array}
\right) \varphi =0\mbox{.}  \label{a1}
\end{equation}

Here \cite{PT}:

$\beta ^{\left[ 1\right] }=\left[ 
\begin{array}{cccc}
0 & 1 & 0 & 0 \\ 
1 & 0 & 0 & 0 \\ 
0 & 0 & 0 & -1 \\ 
0 & 0 & -1 & 0
\end{array}
\right] ,\beta ^{\left[ 2\right] }=\left[ 
\begin{array}{cccc}
0 & -\mathrm{i} & 0 & 0 \\ 
\mathrm{i} & 0 & 0 & 0 \\ 
0 & 0 & 0 & \mathrm{i} \\ 
0 & 0 & -\mathrm{i} & 0
\end{array}
\right]$ ,

$\beta ^{\left[ 3\right] }=\left[ 
\begin{array}{cccc}
1 & 0 & 0 & 0 \\ 
0 & -1 & 0 & 0 \\ 
0 & 0 & -1 & 0 \\ 
0 & 0 & 0 & 1
\end{array}
\right] $

are the diagonal elements of the light Clifford pentad;

$\gamma ^{[0]}=\left[ 
\begin{array}{cccc}
0 & 0 & 1 & 0 \\ 
0 & 0 & 0 & 1 \\ 
1 & 0 & 0 & 0 \\ 
0 & 1 & 0 & 0
\end{array}
\right] ,\beta ^{\left[ 4\right] }=\left[ 
\begin{array}{cccc}
0 & 0 & \mathrm{i} & 0 \\ 
0 & 0 & 0 & \mathrm{i} \\ 
-\mathrm{i} & 0 & 0 & 0 \\ 
0 & -\mathrm{i} & 0 & 0
\end{array}
\right] $

are the antidiagonal (mass) elements of that pentad;

$\Theta _k$ and $\Upsilon _0\gamma ^{\left[ 5\right] }$ with $k\in \left\{
0,1,2,3\right\} $ are fields, forming the electroweak gauge fields \cite{PT2}%
;

$M_0$, $M_4$ are lepton's mass numbers\cite{PT3};

$M_{\zeta ,0}$, $M_{\zeta ,4}$ are red down and up quark's mass numbers;

$M_{\eta ,0}$, $M_{\eta ,4}$ are green down and up quark's mass numbers;

$M_{\theta ,0}$, $M_{\theta ,4}$ are blue down and up quark's mass numbers.

And

$\gamma ^{\left[ 5\right] }=\left[ 
\begin{array}{cccc}
1 & 0 & 0 & 0 \\ 
0 & 1 & 0 & 0 \\ 
0 & 0 & -1 & 0 \\ 
0 & 0 & 0 & -1
\end{array}
\right] $; $\beta ^{\left[ 0\right] }=-\left[ 
\begin{array}{cccc}
1 & 0 & 0 & 0 \\ 
0 & 1 & 0 & 0 \\ 
0 & 0 & 1 & 0 \\ 
0 & 0 & 0 & 1
\end{array}
\right] =-1_4$ \cite{PT1}

The equation (\ref{a1}) can be reformulated as the following \cite{Four}:

\begin{equation}
\frac 1{\mathrm{c}}\partial _t\varphi +\left( \mathrm{i}\Theta _0+\mathrm{i}%
\Upsilon _0\gamma ^{\left[ 5\right] }\right) \varphi =\left( 
\begin{array}{c}
\beta ^{\left[ 1\right] }\left( \partial _1+\mathrm{i}\Theta _1+\mathrm{i}%
\Upsilon _1\gamma ^{\left[ 5\right] }\right) + \\ 
+\beta ^{\left[ 2\right] }\left( \partial _2+\mathrm{i}\Theta _2+\mathrm{i}%
\Upsilon _2\gamma ^{\left[ 5\right] }\right) + \\ 
+\beta ^{\left[ 3\right] }\left( \partial _3+\mathrm{i}\Theta _3+\mathrm{i}%
\Upsilon _3\gamma ^{\left[ 5\right] }\right) + \\ 
+\mathrm{i}M_0\gamma ^{\left[ 0\right] }+\mathrm{i}M_4\beta ^{\left[
4\right] }- \\ 
-\mathrm{i}M_{\zeta ,0}\gamma _\zeta ^{[0]}+\mathrm{i}M_{\zeta ,4}\zeta
^{[4]}- \\ 
-\mathrm{i}M_{\eta ,0}\gamma _\eta ^{[0]}-\mathrm{i}M_{\eta ,4}\eta ^{[4]}+
\\ 
+\mathrm{i}M_{\theta ,0}\gamma _\theta ^{[0]}+\mathrm{i}M_{\theta ,4}\theta
^{[4]}
\end{array}
\right) \varphi \mbox{.}  \label{ham0}
\end{equation}

Here \cite{PT}:

$\gamma _\zeta ^{[0]}=\left[ 
\begin{array}{cccc}
0 & 0 & 0 & -1 \\ 
0 & 0 & -1 & 0 \\ 
0 & -1 & 0 & 0 \\ 
-1 & 0 & 0 & 0
\end{array}
\right] $, $\zeta ^{[4]}=\left[ 
\begin{array}{cccc}
0 & 0 & 0 & \mathrm{i} \\ 
0 & 0 & \mathrm{i} & 0 \\ 
0 & -\mathrm{i} & 0 & 0 \\ 
-\mathrm{i} & 0 & 0 & 0
\end{array}
\right] $

are the antidiagonal (mass) elements of red pentad;

$\gamma _\eta ^{[0]}=\left[ 
\begin{array}{cccc}
0 & 0 & 0 & \mathrm{i} \\ 
0 & 0 & -\mathrm{i} & 0 \\ 
0 & \mathrm{i} & 0 & 0 \\ 
-\mathrm{i} & 0 & 0 & 0
\end{array}
\right] $, $\eta ^{[4]}=\left[ 
\begin{array}{cccc}
0 & 0 & 0 & 1 \\ 
0 & 0 & -1 & 0 \\ 
0 & -1 & 0 & 0 \\ 
1 & 0 & 0 & 0
\end{array}
\right] $

are the antidiagonal (mass) elements of green pentad;

$\gamma _\theta ^{[0]}=\left[ 
\begin{array}{cccc}
0 & 0 & -1 & 0 \\ 
0 & 0 & 0 & 1 \\ 
-1 & 0 & 0 & 0 \\ 
0 & 1 & 0 & 0
\end{array}
\right] $; $\theta ^{[4]}=\left[ 
\begin{array}{cccc}
0 & 0 & -\mathrm{i} & 0 \\ 
0 & 0 & 0 & \mathrm{i} \\ 
-\mathrm{i} & 0 & 0 & 0 \\ 
0 & \mathrm{i} & 0 & 0
\end{array}
\right] $

are the antidiagonal (mass) elements of blue pentad.

The quarks mass members of this equation form the following matrix sum:

\[
\widehat{M}\stackrel{def}{=} 
\begin{array}{c}
-M_{\zeta ,0}\gamma _\zeta ^{[0]}+M_{\zeta ,4}\zeta ^{[4]}- \\ 
-M_{\eta ,0}\gamma _\eta ^{[0]}-M_{\eta ,4}\eta ^{[4]}+ \\ 
+M_{\theta ,0}\gamma _\theta ^{[0]}+M_{\theta ,4}\theta ^{[4]}
\end{array}
\]

\[
=\left[ 
\begin{array}{cccc}
0 & 0 & -M_{\theta ,0} & M_{\zeta ,0}-\mathrm{i}M_{\eta ,0} \\ 
0 & 0 & M_{\zeta ,0}+\mathrm{i}M_{\eta ,0} & M_{\theta ,0} \\ 
-M_{\theta ,0} & M_{\zeta ,0}-\mathrm{i}M_{\eta ,0} & 0 & 0 \\ 
M_{\zeta ,0}+\mathrm{i}M_{\eta ,0} & M_{\theta ,0} & 0 & 0
\end{array}
\right] 
\]

\[
+\mathrm{i}\left[ 
\begin{array}{cccc}
0 & 0 & -M_{\theta ,4} & M_{\zeta ,4}+\mathrm{i}M_{\eta ,4} \\ 
0 & 0 & M_{\zeta ,4}-\mathrm{i}M_{\eta ,4} & M_{\theta ,4} \\ 
-M_{\theta ,4} & -M_{\zeta ,4}-\mathrm{i}M_{\eta ,4} & 0 & 0 \\ 
-M_{\zeta ,4}+\mathrm{i}M_{\eta ,4} & M_{\theta ,4} & 0 & 0
\end{array}
\right] \mbox{.} 
\]

Elements of these matrices can be turned by formula of shape \cite{Z}:

\[
=\left( 
\begin{array}{cc}
\cos \frac \theta 2 & \mathrm{i}\sin \frac \theta 2 \\ 
\mathrm{i}\sin \frac \theta 2 & \cos \frac \theta 2
\end{array}
\right) \left( 
\begin{array}{cc}
Z & X-\mathrm{i}Y \\ 
X+\mathrm{i}Y & -Z
\end{array}
\right) \left( 
\begin{array}{cc}
\cos \frac \theta 2 & -\mathrm{i}\sin \frac \theta 2 \\ 
-\mathrm{i}\sin \frac \theta 2 & \cos \frac \theta 2
\end{array}
\right) 
\]

\[
=\left( 
\begin{array}{cc}
Z\cos \theta -Y\sin \theta & X-\mathrm{i}\left( Y\cos \theta +Z\sin \theta
\right) \\ 
X+\mathrm{i}\left( Y\cos \theta +Z\sin \theta \right) & -Z\cos \theta +Y\sin
\theta
\end{array}
\right) \mbox{.} 
\]

Hence, if:

\[
U_{2,3}\left( \alpha \right) \stackrel{def}{=}\left[ 
\begin{array}{cccc}
\cos \alpha & \mathrm{i}\sin \alpha & 0 & 0 \\ 
\mathrm{i}\sin \alpha & \cos \alpha & 0 & 0 \\ 
0 & 0 & \cos \alpha & \mathrm{i}\sin \alpha \\ 
0 & 0 & \mathrm{i}\sin \alpha & \cos \alpha
\end{array}
\right] 
\]

and

\[
\widehat{M}^{\prime }\stackrel{def}{=} 
\begin{array}{c}
-M_{\zeta ,0}^{\prime }\gamma _\zeta ^{[0]}+M_{\zeta ,4}^{\prime }\zeta
^{[4]}- \\ 
-M_{\eta ,0}^{\prime }\gamma _\eta ^{[0]}-M_{\eta ,4}^{\prime }\eta ^{[4]}+
\\ 
+M_{\theta ,0}^{\prime }\gamma _\theta ^{[0]}+M_{\theta ,4}^{\prime }\theta
^{[4]}
\end{array}
\stackrel{def}{=}U_{2,3}^{\dagger }\left( \alpha \right) \widehat{M}%
U_{2,3}\left( \alpha \right) 
\]

then:

$M_{\zeta ,0}^{\prime }=M_{\zeta ,0}$,

$M_{\eta ,0}^{\prime }=M_{\eta ,0}\cos 2\alpha +M_{\theta ,0}\sin 2\alpha $,

$M_{\theta ,0}^{\prime }=M_{\theta ,0}\cos 2\alpha -M_{\eta ,0}\sin 2\alpha $%
,

$M_{\zeta ,4}^{\prime }=M_{\zeta ,4}$,

$M_{\eta ,4}^{\prime }=M_{\eta ,4}\cos 2\alpha +M_{\theta ,4}\sin 2\alpha $,

$M_{\theta ,4}^{\prime }=M_{\theta ,4}\cos 2\alpha -M_{\eta ,4}\sin 2\alpha $%
,

Therefore, matrix $U_{2,3}\left( \alpha \right) $ makes an oscillation
between green and blue quarks colors.

Let us consider equation (\ref{ham0}) under transformation $U_{2,3}\left(
\alpha \right) $ with $\alpha $ is an arbitrary real function of time-space
variables ($\alpha =\alpha \left( t,x_1,x_2,x_3\right) $):

\[
U_{2,3}^{\dagger }\left( \alpha \right) \left( \frac 1{\mathrm{c}}\partial
_t+\mathrm{i}\Theta _0+\mathrm{i}\Upsilon _0\gamma ^{\left[ 5\right]
}\right) U_{2,3}\left( \alpha \right) \varphi = 
\]

\[
=U_{2,3}^{\dagger }\left( \alpha \right) \left( 
\begin{array}{c}
\beta ^{\left[ 1\right] }\left( \partial _1+\mathrm{i}\Theta _1+\mathrm{i}%
\Upsilon _1\gamma ^{\left[ 5\right] }\right) + \\ 
+\beta ^{\left[ 2\right] }\left( \partial _2+\mathrm{i}\Theta _2+\mathrm{i}%
\Upsilon _2\gamma ^{\left[ 5\right] }\right) + \\ 
+\beta ^{\left[ 3\right] }\left( \partial _3+\mathrm{i}\Theta _3+\mathrm{i}%
\Upsilon _3\gamma ^{\left[ 5\right] }\right) + \\ 
+\mathrm{i}M_0\gamma ^{\left[ 0\right] }+\mathrm{i}M_4\beta ^{\left[
4\right] }+\widehat{M}
\end{array}
\right) U_{2,3}\left( \alpha \right) \varphi \mbox{.} 
\]

That is:

\[
\left( 
\begin{array}{c}
\frac 1{\mathrm{c}}U_{2,3}^{\dagger }\left( \alpha \right) \partial _t\left(
U_{2,3}\left( \alpha \right) \varphi \right) \\ 
+\mathrm{i}\Theta _0U_{2,3}^{\dagger }\left( \alpha \right) \left(
U_{2,3}\left( \alpha \right) \varphi \right) \\ 
+\mathrm{i}\Upsilon _0U_{2,3}^{\dagger }\left( \alpha \right) \gamma
^{\left[ 5\right] }\left( U_{2,3}\left( \alpha \right) \varphi \right)
\end{array}
\right) = 
\]

\[
=U_{2,3}^{\dagger }\left( \alpha \right) \left( 
\begin{array}{c}
\beta ^{\left[ 1\right] }\left( \partial _1+\mathrm{i}\Theta _1+\mathrm{i}%
\Upsilon _1\gamma ^{\left[ 5\right] }\right) + \\ 
+\beta ^{\left[ 2\right] }\left( \partial _2+\mathrm{i}\Theta _2+\mathrm{i}%
\Upsilon _2\gamma ^{\left[ 5\right] }\right) + \\ 
+\beta ^{\left[ 3\right] }\left( \partial _3+\mathrm{i}\Theta _3+\mathrm{i}%
\Upsilon _3\gamma ^{\left[ 5\right] }\right) + \\ 
+\mathrm{i}M_0\gamma ^{\left[ 0\right] }+\mathrm{i}M_4\beta ^{\left[
4\right] }+\widehat{M}
\end{array}
\right) U_{2,3}\left( \alpha \right) \varphi \mbox{.} 
\]

Because

\[
\begin{array}{c}
U_{2,3}^{\dagger }\left( \alpha \right) U_{2,3}\left( \alpha \right) =1_4%
\mbox{,} \\ 
U_{2,3}^{\dagger }\left( \alpha \right) \gamma ^{\left[ 5\right]
}U_{2,3}\left( \alpha \right) =\gamma ^{\left[ 5\right] }
\end{array}
\]

then

\[
\left( 
\begin{array}{c}
\frac 1{\mathrm{c}}\left( \left( U_{2,3}^{\dagger }\left( \alpha \right)
\partial _tU_{2,3}\left( \alpha \right) \right) \varphi +U_{2,3}^{\dagger
}\left( \alpha \right) U_{2,3}\left( \alpha \right) \partial _t\varphi
\right) \\ 
+\mathrm{i}\Theta _0\varphi +\mathrm{i}\Upsilon _0\gamma ^{\left[ 5\right]
}\varphi
\end{array}
\right) = 
\]

\[
=\left( 
\begin{array}{c}
U_{2,3}^{\dagger }\left( \alpha \right) \beta ^{\left[ 1\right] }\left( 
\begin{array}{c}
\left( \partial _1U_{2,3}\left( \alpha \right) \right) \varphi
+U_{2,3}\left( \alpha \right) \partial _1\varphi \\ 
+\mathrm{i}\Theta _1\left( U_{2,3}\left( \alpha \right) \varphi \right) +%
\mathrm{i}\Upsilon _1\gamma ^{\left[ 5\right] }\left( U_{2,3}\left( \alpha
\right) \varphi \right)
\end{array}
\right) + \\ 
+U_{2,3}^{\dagger }\left( \alpha \right) \beta ^{\left[ 2\right] }\left( 
\begin{array}{c}
\left( \partial _2U_{2,3}\left( \alpha \right) \right) \varphi
+U_{2,3}\left( \alpha \right) \partial _2\varphi \\ 
+\mathrm{i}\Theta _2\left( U_{2,3}\left( \alpha \right) \varphi \right) +%
\mathrm{i}\Upsilon _2\gamma ^{\left[ 5\right] }\left( U_{2,3}\left( \alpha
\right) \varphi \right)
\end{array}
\right) + \\ 
+U_{2,3}^{\dagger }\left( \alpha \right) \beta ^{\left[ 3\right] }\left( 
\begin{array}{c}
\left( \partial _3U_{2,3}\left( \alpha \right) \right) \varphi
+U_{2,3}\left( \alpha \right) \partial _3\varphi \\ 
+\mathrm{i}\Theta _3\left( U_{2,3}\left( \alpha \right) \varphi \right) +%
\mathrm{i}\Upsilon _3\gamma ^{\left[ 5\right] }\left( U_{2,3}\left( \alpha
\right) \varphi \right)
\end{array}
\right) + \\ 
+\mathrm{i}M_0U_{2,3}^{\dagger }\left( \alpha \right) \gamma ^{\left[
0\right] }\left( U_{2,3}\left( \alpha \right) \varphi \right) \\ 
+\mathrm{i}M_4U_{2,3}^{\dagger }\left( \alpha \right) \beta ^{\left[
4\right] }\left( U_{2,3}\left( \alpha \right) \varphi \right) \\ 
+U_{2,3}^{\dagger }\left( \alpha \right) \widehat{M}\left( U_{2,3}\left(
\alpha \right) \varphi \right)
\end{array}
\right) \mbox{.} 
\]

Since

\[
\begin{array}{c}
U_{2,3}^{\dagger }\left( \alpha \right) \gamma ^{\left[ 0\right]
}U_{2,3}\left( \alpha \right) =\gamma ^{\left[ 0\right] }\mbox{,} \\ 
U_{2,3}^{\dagger }\left( \alpha \right) \beta ^{\left[ 4\right]
}U_{2,3}\left( \alpha \right) =\beta ^{\left[ 4\right] }
\end{array}
\]

then

\[
\left( 
\begin{array}{c}
\frac 1{\mathrm{c}}\left( \left( U_{2,3}^{\dagger }\left( \alpha \right)
\partial _tU_{2,3}\left( \alpha \right) \right) \varphi +U_{2,3}^{\dagger
}\left( \alpha \right) U_{2,3}\left( \alpha \right) \partial _t\varphi
\right) \\ 
+\mathrm{i}\Theta _0\varphi +\mathrm{i}\Upsilon _0\gamma ^{\left[ 5\right]
}\varphi
\end{array}
\right) = 
\]

\[
=\left( 
\begin{array}{c}
U_{2,3}^{\dagger }\left( \alpha \right) \beta ^{\left[ 1\right] }\left( 
\begin{array}{c}
\left( \partial _1U_{2,3}\left( \alpha \right) \right) \varphi
+U_{2,3}\left( \alpha \right) \partial _1\varphi \\ 
+\mathrm{i}\Theta _1\left( U_{2,3}\left( \alpha \right) \varphi \right) +%
\mathrm{i}\Upsilon _1\gamma ^{\left[ 5\right] }\left( U_{2,3}\left( \alpha
\right) \varphi \right)
\end{array}
\right) + \\ 
+U_{2,3}^{\dagger }\left( \alpha \right) \beta ^{\left[ 2\right] }\left( 
\begin{array}{c}
\left( \partial _2U_{2,3}\left( \alpha \right) \right) \varphi
+U_{2,3}\left( \alpha \right) \partial _2\varphi \\ 
+\mathrm{i}\Theta _2\left( U_{2,3}\left( \alpha \right) \varphi \right) +%
\mathrm{i}\Upsilon _2\gamma ^{\left[ 5\right] }\left( U_{2,3}\left( \alpha
\right) \varphi \right)
\end{array}
\right) + \\ 
+U_{2,3}^{\dagger }\left( \alpha \right) \beta ^{\left[ 3\right] }\left( 
\begin{array}{c}
\left( \partial _3U_{2,3}\left( \alpha \right) \right) \varphi
+U_{2,3}\left( \alpha \right) \partial _3\varphi \\ 
+\mathrm{i}\Theta _3\left( U_{2,3}\left( \alpha \right) \varphi \right) +%
\mathrm{i}\Upsilon _3\gamma ^{\left[ 5\right] }\left( U_{2,3}\left( \alpha
\right) \varphi \right)
\end{array}
\right) + \\ 
+\mathrm{i}M_0\gamma ^{\left[ 0\right] }\varphi +\mathrm{i}M_4\beta ^{\left[
4\right] }\varphi +\widehat{M}^{\prime }\varphi
\end{array}
\right) \mbox{.} 
\]

Since

\[
U_{2,3}^{\dagger }\left( \alpha \right) \beta ^{\left[ 1\right] }=\beta
^{\left[ 1\right] }U_{2,3}^{\dagger }\left( \alpha \right) 
\]

then

\[
\left( U_{2,3}^{\dagger }\left( \alpha \right) \frac 1{\mathrm{c}}\partial
_tU_{2,3}\left( \alpha \right) +\frac 1{\mathrm{c}}\partial _t+\mathrm{i}%
\Theta _0+\mathrm{i}\Upsilon _0\gamma ^{\left[ 5\right] }\right) \varphi = 
\]

\[
=\left( 
\begin{array}{c}
\beta ^{\left[ 1\right] }U_{2,3}^{\dagger }\left( \alpha \right) \left( 
\begin{array}{c}
\left( \partial _1U_{2,3}\left( \alpha \right) \right) +U_{2,3}\left( \alpha
\right) \partial _1 \\ 
+\mathrm{i}\Theta _1U_{2,3}\left( \alpha \right) +\mathrm{i}\Upsilon
_1\gamma ^{\left[ 5\right] }U_{2,3}\left( \alpha \right)
\end{array}
\right) + \\ 
+U_{2,3}^{\dagger }\left( \alpha \right) \beta ^{\left[ 2\right] }\left( 
\begin{array}{c}
\left( \partial _2U_{2,3}\left( \alpha \right) \right) +U_{2,3}\left( \alpha
\right) \partial _2 \\ 
+\mathrm{i}\Theta _2U_{2,3}\left( \alpha \right) +\mathrm{i}\Upsilon
_2\gamma ^{\left[ 5\right] }U_{2,3}\left( \alpha \right)
\end{array}
\right) + \\ 
+U_{2,3}^{\dagger }\left( \alpha \right) \beta ^{\left[ 3\right] }\left( 
\begin{array}{c}
\left( \partial _3U_{2,3}\left( \alpha \right) \right) +U_{2,3}\left( \alpha
\right) \partial _3 \\ 
+\mathrm{i}\Theta _3U_{2,3}\left( \alpha \right) +\mathrm{i}\Upsilon
_3\gamma ^{\left[ 5\right] }U_{2,3}\left( \alpha \right)
\end{array}
\right) + \\ 
+\mathrm{i}M_0\gamma ^{\left[ 0\right] }+\mathrm{i}M_4\beta ^{\left[
4\right] }+\widehat{M}^{\prime }
\end{array}
\right) \varphi \mbox{.} 
\]

Because

\[
\begin{array}{c}
U_{2,3}^{\dagger }\left( \alpha \right) \beta ^{\left[ 2\right] }=\left(
\beta ^{\left[ 2\right] }\cos 2\alpha +\beta ^{\left[ 3\right] }\sin 2\alpha
\right) U_{2,3}^{\dagger }\left( \alpha \right) \mbox{,} \\ 
U_{2,3}^{\dagger }\left( \alpha \right) \beta ^{\left[ 3\right] }=\left(
\beta ^{\left[ 3\right] }\cos 2\alpha -\beta ^{\left[ 2\right] }\sin 2\alpha
\right) U_{2,3}^{\dagger }\left( \alpha \right)
\end{array}
\]

then

\[
\left( U_{2,3}^{\dagger }\left( \alpha \right) \frac 1{\mathrm{c}}\partial
_tU_{2,3}\left( \alpha \right) +\frac 1{\mathrm{c}}\partial _t+\mathrm{i}%
\Theta _0+\mathrm{i}\Upsilon _0\gamma ^{\left[ 5\right] }\right) \varphi = 
\]

\[
=\left( 
\begin{array}{c}
\beta ^{\left[ 1\right] }\left( 
\begin{array}{c}
U_{2,3}^{\dagger }\left( \alpha \right) \partial _1U_{2,3}\left( \alpha
\right) +U_{2,3}^{\dagger }\left( \alpha \right) U_{2,3}\left( \alpha
\right) \partial _1 \\ 
+\mathrm{i}\Theta _1U_{2,3}^{\dagger }\left( \alpha \right) U_{2,3}\left(
\alpha \right) +\mathrm{i}\Upsilon _1U_{2,3}^{\dagger }\left( \alpha \right)
\gamma ^{\left[ 5\right] }U_{2,3}\left( \alpha \right)
\end{array}
\right) + \\ 
+\left( \beta ^{\left[ 2\right] }\cos 2\alpha +\beta ^{\left[ 3\right] }\sin
2\alpha \right) \\ 
\left( 
\begin{array}{c}
U_{2,3}^{\dagger }\left( \alpha \right) \partial _2U_{2,3}\left( \alpha
\right) +U_{2,3}^{\dagger }\left( \alpha \right) U_{2,3}\left( \alpha
\right) \partial _2 \\ 
+\mathrm{i}\Theta _2U_{2,3}^{\dagger }\left( \alpha \right) U_{2,3}\left(
\alpha \right) +\mathrm{i}\Upsilon _2U_{2,3}^{\dagger }\left( \alpha \right)
\gamma ^{\left[ 5\right] }U_{2,3}\left( \alpha \right)
\end{array}
\right) + \\ 
+\left( \beta ^{\left[ 3\right] }\cos 2\alpha -\beta ^{\left[ 2\right] }\sin
2\alpha \right) \\ 
\left( 
\begin{array}{c}
U_{2,3}^{\dagger }\left( \alpha \right) \partial _3U_{2,3}\left( \alpha
\right) +U_{2,3}^{\dagger }\left( \alpha \right) U_{2,3}\left( \alpha
\right) \partial _3 \\ 
+\mathrm{i}\Theta _3U_{2,3}^{\dagger }\left( \alpha \right) U_{2,3}\left(
\alpha \right) +\mathrm{i}\Upsilon _3U_{2,3}^{\dagger }\left( \alpha \right)
\gamma ^{\left[ 5\right] }U_{2,3}\left( \alpha \right)
\end{array}
\right) + \\ 
+\mathrm{i}M_0\gamma ^{\left[ 0\right] }+\mathrm{i}M_4\beta ^{\left[
4\right] }+\widehat{M}^{\prime }
\end{array}
\right) \varphi \mbox{.} 
\]

Hence

\[
\left( \frac 1{\mathrm{c}}\partial _t+U_{2,3}^{\dagger }\left( \alpha
\right) \frac 1{\mathrm{c}}\partial _tU_{2,3}\left( \alpha \right) +\mathrm{i%
}\Theta _0+\mathrm{i}\Upsilon _0\gamma ^{\left[ 5\right] }\right) \varphi = 
\]

\begin{equation}
=\left( 
\begin{array}{c}
\beta ^{\left[ 1\right] }\left( \partial _1+U_{2,3}^{\dagger }\left( \alpha
\right) \partial _1U_{2,3}\left( \alpha \right) +\mathrm{i}\Theta _1+\mathrm{%
i}\Upsilon _1\gamma ^{\left[ 5\right] }\right) \\ 
+\beta ^{\left[ 2\right] }\left( 
\begin{array}{c}
\left( \cos 2\alpha \cdot \partial _2-\sin 2\alpha \cdot \partial _3\right)
\\ 
+U_{2,3}^{\dagger }\left( \alpha \right) \left( \cos 2\alpha \cdot \partial
_2-\sin 2\alpha \cdot \partial _3\right) U_{2,3}\left( \alpha \right) \\ 
+\mathrm{i}\left( \Theta _2\cos 2\alpha -\Theta _3\sin 2\alpha \right) \\ 
+\mathrm{i}\left( \Upsilon _2\gamma ^{\left[ 5\right] }\cos 2\alpha
-\Upsilon _3\gamma ^{\left[ 5\right] }\sin 2\alpha \right)
\end{array}
\right) \\ 
+\beta ^{\left[ 3\right] }\left( 
\begin{array}{c}
\left( \cos 2\alpha \cdot \partial _3+\sin 2\alpha \cdot \partial _2\right)
\\ 
+U_{2,3}^{\dagger }\left( \alpha \right) \left( \cos 2\alpha \cdot \partial
_3+\sin 2\alpha \cdot \partial _2\right) U_{2,3}\left( \alpha \right) \\ 
+\mathrm{i}\left( \Theta _2\sin 2\alpha +\Theta _3\cos 2\alpha \right) \\ 
+\mathrm{i}\left( \Upsilon _3\gamma ^{\left[ 5\right] }\cos 2\alpha
+\Upsilon _2\gamma ^{\left[ 5\right] }\sin 2\alpha \right)
\end{array}
\right) + \\ 
+\mathrm{i}M_0\gamma ^{\left[ 0\right] }+\mathrm{i}M_4\beta ^{\left[
4\right] }+\widehat{M}^{\prime }
\end{array}
\right) \varphi \mbox{.}  \label{ham5}
\end{equation}

Let $x_2^{\prime }$ and $x_3^{\prime }$ are elements of other coordinates
system such that:

\begin{eqnarray*}
&&\frac{\partial x_2}{\partial x_2^{\prime }}=\cos 2\alpha \mbox{,} \\
&&\frac{\partial x_3}{\partial x_2^{\prime }}=-\sin 2\alpha \mbox{,} \\
&&\frac{\partial x_2}{\partial x_3^{\prime }}=\sin 2\alpha \mbox{.} \\
&&\frac{\partial x_3}{\partial x_3^{\prime }}=\cos 2\alpha \mbox{,} \\
&&\frac{\partial x_0}{\partial x_2^{\prime }}=\frac{\partial x_1}{\partial
x_2^{\prime }}=\frac{\partial x_0}{\partial x_3^{\prime }}=\frac{\partial x_1%
}{\partial x_3^{\prime }}=0\mbox{.}
\end{eqnarray*}

Hence:

\begin{eqnarray*}
\partial _2^{\prime }\stackrel{def}{=}\frac \partial {\partial x_2^{\prime
}} &=&\frac \partial {\partial x_0}\frac{\partial x_0}{\partial x_2^{\prime }%
}+\frac \partial {\partial x_1}\frac{\partial x_1}{\partial x_2^{\prime }}%
+\frac \partial {\partial x_2}\frac{\partial x_2}{\partial x_2^{\prime }}%
+\frac \partial {\partial x_3}\frac{\partial x_3}{\partial x_2^{\prime }} \\
&=&\cos 2\alpha \cdot \frac \partial {\partial x_2}-\sin 2\alpha \cdot \frac
\partial {\partial x_3} \\
&=&\cos 2\alpha \cdot \partial _2-\sin 2\alpha \cdot \partial _3\mbox{,}
\end{eqnarray*}

\begin{eqnarray*}
\partial _3^{\prime }\stackrel{def}{=}\frac \partial {\partial x_3^{\prime
}} &=&\frac \partial {\partial x_0}\frac{\partial x_0}{\partial x_3^{\prime }%
}+\frac \partial {\partial x_1}\frac{\partial x_1}{\partial x_3^{\prime }}%
+\frac \partial {\partial x_2}\frac{\partial x_2}{\partial x_3^{\prime }}%
+\frac \partial {\partial x_3}\frac{\partial x_3}{\partial x_3^{\prime }} \\
&=&\cos 2\alpha \cdot \frac \partial {\partial x_3}+\sin 2\alpha \cdot \frac
\partial {\partial x_2} \\
&=&\cos 2\alpha \cdot \partial _3+\sin 2\alpha \cdot \partial _2.
\end{eqnarray*}

Therefore from (\ref{ham5}):

\[
\left( \frac 1{\mathrm{c}}\partial _t+U_{2,3}^{\dagger }\left( \alpha
\right) \frac 1{\mathrm{c}}\partial _tU_{2,3}\left( \alpha \right) +\mathrm{i%
}\Theta _0+\mathrm{i}\Upsilon _0\gamma ^{\left[ 5\right] }\right) \varphi = 
\]

\[
=\left( 
\begin{array}{c}
\beta ^{\left[ 1\right] }\left( \partial _1+U_{2,3}^{\dagger }\left( \alpha
\right) \partial _1U_{2,3}\left( \alpha \right) +\mathrm{i}\Theta _1+\mathrm{%
i}\Upsilon _1\gamma ^{\left[ 5\right] }\right) \\ 
+\beta ^{\left[ 2\right] }\left( \partial _2^{\prime }+U_{2,3}^{\dagger
}\left( \alpha \right) \partial _2^{\prime }U_{2,3}\left( \alpha \right) +%
\mathrm{i}\Theta _2^{\prime }+\mathrm{i}\Upsilon _2^{\prime }\gamma ^{\left[
5\right] }\right) \\ 
+\beta ^{\left[ 3\right] }\left( \partial _3^{\prime }+U_{2,3}^{\dagger
}\left( \alpha \right) \partial _3^{\prime }U_{2,3}\left( \alpha \right) +%
\mathrm{i}\Theta _3^{\prime }+\mathrm{i}\Upsilon _3^{\prime }\gamma ^{\left[
5\right] }\right) + \\ 
+\mathrm{i}M_0\gamma ^{\left[ 0\right] }+\mathrm{i}M_4\beta ^{\left[
4\right] }+\widehat{M}^{\prime }
\end{array}
\right) \varphi \mbox{.} 
\]

with

\[
\begin{array}{c}
\Theta _2^{\prime }\stackrel{def}{=}\Theta _2\cos 2\alpha -\Theta _3\sin
2\alpha \mbox{,} \\ 
\Theta _3^{\prime }\stackrel{def}{=}\Theta _2\sin 2\alpha +\Theta _3\cos
2\alpha \mbox{.} \\ 
\Upsilon _2^{\prime }\stackrel{def}{=}\Upsilon _2\cos 2\alpha -\Upsilon
_3\sin 2\alpha \mbox{,} \\ 
\Upsilon _3^{\prime }\stackrel{def}{=}\Upsilon _3\cos 2\alpha +\Upsilon
_2\sin 2\alpha \mbox{.}
\end{array}
\]

Therefore, the oscillation between blue and green quarks colors curves the
space of events in the $x_2$, $x_3$ directions.

Similarly that: matrix

\[
U_{1,3}\left( \vartheta \right) \stackrel{def}{=}\left[ 
\begin{array}{cccc}
\cos \vartheta & \sin \vartheta & 0 & 0 \\ 
-\sin \vartheta & \cos \vartheta & 0 & 0 \\ 
0 & 0 & \cos \vartheta & \sin \vartheta \\ 
0 & 0 & -\sin \vartheta & \cos \vartheta
\end{array}
\right] 
\]

with an arbitrary real function $\vartheta \left( t,x_1,x_2,x_3\right) $
describes the oscillation between blue and red quarks colors which curves
the space of events in the $x_1$, $x_3$ directions. And matrix

\[
U_{1,2}\left( \varsigma \right) \stackrel{def}{=}\left[ 
\begin{array}{cccc}
\cos \varsigma -i\sin \varsigma & 0 & 0 & 0 \\ 
0 & \cos \varsigma +i\sin \varsigma & 0 & 0 \\ 
0 & 0 & \cos \varsigma -i\sin \varsigma & 0 \\ 
0 & 0 & 0 & \cos \varsigma +i\sin \varsigma
\end{array}
\right] 
\]

with an arbitrary real function $\varsigma \left( t,x_1,x_2,x_3\right) $
describes the oscillation between green and red quarks colors which curves
the space of events in the $x_1$, $x_2$ directions.

Now, let

\[
U_{0,1}\left( \upsilon \right) \stackrel{def}{=}\left[ 
\begin{array}{cccc}
\cosh \upsilon  & -\sinh \upsilon  & 0 & 0 \\ 
-\sinh \upsilon  & \cosh \upsilon  & 0 & 0 \\ 
0 & 0 & \cosh \upsilon  & \sinh \upsilon  \\ 
0 & 0 & \sinh \upsilon  & \cosh \upsilon 
\end{array}
\right] \mbox{.}
\]

and

\[
\widehat{M}^{\prime \prime }\stackrel{def}{=}
\begin{array}{c}
-M_{\zeta ,0}^{\prime \prime }\gamma _\zeta ^{[0]}+M_{\zeta ,4}^{\prime
\prime }\zeta ^{[4]}- \\ 
-M_{\eta ,0}^{\prime \prime }\gamma _\eta ^{[0]}-M_{\eta ,4}^{\prime \prime
}\eta ^{[4]}+ \\ 
+M_{\theta ,0}^{\prime \prime }\gamma _\theta ^{[0]}+M_{\theta ,4}^{\prime
\prime }\theta ^{[4]}
\end{array}
\stackrel{def}{=}U_{0,1}^{\dagger }\left( \upsilon \right) \widehat{M}%
U_{0,1}\left( \upsilon \right) 
\]

then:

$M_{\zeta ,0}^{\prime \prime }=M_{\zeta ,0}$,

$M_{\eta ,0}^{\prime \prime }=M_{\eta ,0}\cosh 2\upsilon -M_{\theta ,4}\sinh
2\upsilon $,

$M_{\theta ,0}^{\prime \prime }=M_{\theta ,0}\cosh 2\upsilon +M_{\eta
,4}\sinh 2\upsilon $,

$M_{\zeta ,4}^{\prime \prime }=M_{\zeta ,4}$

$M_{\eta ,4}^{\prime \prime }=M_{\eta ,4}\cosh 2\upsilon +M_{\theta ,0}\sinh
2\upsilon $,

$M_{\theta ,4}^{\prime \prime }=M_{\theta ,4}\cosh 2\upsilon -M_{\eta
,0}\sinh 2\upsilon $,

Therefore, matrix $U_{0,1}\left( \upsilon \right) $ makes an oscillation
between green and blue quarks colors with an oscillation between up and down
quarks.

Let us consider equation (\ref{ham0}) under transformation $U_{0,1}\left(
\upsilon \right) $ with $\upsilon $ is an arbitrary real function of
time-space variables ($\upsilon =\upsilon \left( t,x_1,x_2,x_3\right) $):

\[
U_{0,1}^{\dagger }\left( \upsilon \right) \left( \frac 1{\mathrm{c}}\partial
_t+\mathrm{i}\Theta _0+\mathrm{i}\Upsilon _0\gamma ^{\left[ 5\right]
}\right) U_{0,1}\left( \upsilon \right) \varphi = 
\]

\[
=U_{0,1}^{\dagger }\left( \upsilon \right) \left( 
\begin{array}{c}
\beta ^{\left[ 1\right] }\left( \partial _1+\mathrm{i}\Theta _1+\mathrm{i}%
\Upsilon _1\gamma ^{\left[ 5\right] }\right) + \\ 
+\beta ^{\left[ 2\right] }\left( \partial _2+\mathrm{i}\Theta _2+\mathrm{i}%
\Upsilon _2\gamma ^{\left[ 5\right] }\right) + \\ 
+\beta ^{\left[ 3\right] }\left( \partial _3+\mathrm{i}\Theta _3+\mathrm{i}%
\Upsilon _3\gamma ^{\left[ 5\right] }\right) + \\ 
+\mathrm{i}M_0\gamma ^{\left[ 0\right] }+\mathrm{i}M_4\beta ^{\left[
4\right] }+\widehat{M}
\end{array}
\right) U_{0,1}\left( \upsilon \right) \varphi \mbox{.} 
\]

Hence,

\[
U_{0,1}^{\dagger }\left( \upsilon \right) \left( 
\begin{array}{c}
\frac 1{\mathrm{c}}\partial _t\left( U_{0,1}\left( \upsilon \right) \varphi
\right) \\ 
+\mathrm{i}\Theta _0\left( U_{0,1}\left( \upsilon \right) \varphi \right) +%
\mathrm{i}\Upsilon _0\gamma ^{\left[ 5\right] }\left( U_{0,1}\left( \upsilon
\right) \varphi \right)
\end{array}
\right) = 
\]

\[
=U_{0,1}^{\dagger }\left( \upsilon \right) \left( 
\begin{array}{c}
\beta ^{\left[ 1\right] }\left( 
\begin{array}{c}
\partial _1\left( U_{0,1}\left( \upsilon \right) \varphi \right) \\ 
+\mathrm{i}\Theta _1\left( U_{0,1}\left( \upsilon \right) \varphi \right) +%
\mathrm{i}\Upsilon _1\gamma ^{\left[ 5\right] }\left( U_{0,1}\left( \upsilon
\right) \varphi \right)
\end{array}
\right) + \\ 
+\beta ^{\left[ 2\right] }\left( 
\begin{array}{c}
\partial _2\left( U_{0,1}\left( \upsilon \right) \varphi \right) \\ 
+\mathrm{i}\Theta _2\left( U_{0,1}\left( \upsilon \right) \varphi \right) +%
\mathrm{i}\Upsilon _2\gamma ^{\left[ 5\right] }\left( U_{0,1}\left( \upsilon
\right) \varphi \right)
\end{array}
\right) + \\ 
+\beta ^{\left[ 3\right] }\left( 
\begin{array}{c}
\partial _3\left( U_{0,1}\left( \upsilon \right) \varphi \right) \\ 
+\mathrm{i}\Theta _3\left( U_{0,1}\left( \upsilon \right) \varphi \right) +%
\mathrm{i}\Upsilon _3\gamma ^{\left[ 5\right] }\left( U_{0,1}\left( \upsilon
\right) \varphi \right)
\end{array}
\right) + \\ 
+\mathrm{i}M_0\gamma ^{\left[ 0\right] }\left( U_{0,1}\left( \upsilon
\right) \varphi \right) \\ 
+\mathrm{i}M_4\beta ^{\left[ 4\right] }\left( U_{0,1}\left( \upsilon \right)
\varphi \right) +\widehat{M}\left( U_{0,1}\left( \upsilon \right) \varphi
\right)
\end{array}
\right) \mbox{.} 
\]

\[
U_{0,1}^{\dagger }\left( \upsilon \right) \left( 
\begin{array}{c}
\frac 1{\mathrm{c}}\left( \partial _tU_{0,1}\left( \upsilon \right) \right)
\varphi +U_{0,1}\left( \upsilon \right) \frac 1{\mathrm{c}}\partial _t\varphi
\\ 
+\mathrm{i}\Theta _0U_{0,1}\left( \upsilon \right) \varphi +\mathrm{i}%
\Upsilon _0\gamma ^{\left[ 5\right] }U_{0,1}\left( \upsilon \right) \varphi
\end{array}
\right) = 
\]

\[
=U_{0,1}^{\dagger }\left( \upsilon \right) \left( 
\begin{array}{c}
\beta ^{\left[ 1\right] }\left( 
\begin{array}{c}
\left( \partial _1U_{0,1}\left( \upsilon \right) \right) \varphi
+U_{0,1}\left( \upsilon \right) \partial _1\varphi \\ 
+\mathrm{i}\Theta _1U_{0,1}\left( \upsilon \right) \varphi +\mathrm{i}%
\Upsilon _1\gamma ^{\left[ 5\right] }U_{0,1}\left( \upsilon \right) \varphi
\end{array}
\right) + \\ 
+\beta ^{\left[ 2\right] }\left( 
\begin{array}{c}
\left( \partial _2U_{0,1}\left( \upsilon \right) \right) \varphi
+U_{0,1}\left( \upsilon \right) \partial _2\varphi \\ 
+\mathrm{i}\Theta _2U_{0,1}\left( \upsilon \right) \varphi +\mathrm{i}%
\Upsilon _2\gamma ^{\left[ 5\right] }U_{0,1}\left( \upsilon \right) \varphi
\end{array}
\right) + \\ 
+\beta ^{\left[ 3\right] }\left( 
\begin{array}{c}
\left( \partial _3U_{0,1}\left( \upsilon \right) \right) \varphi
+U_{0,1}\left( \upsilon \right) \partial _3\varphi \\ 
+\mathrm{i}\Theta _3U_{0,1}\left( \upsilon \right) \varphi +\mathrm{i}%
\Upsilon _3\gamma ^{\left[ 5\right] }U_{0,1}\left( \upsilon \right) \varphi
\end{array}
\right) + \\ 
+\mathrm{i}M_0\gamma ^{\left[ 0\right] }U_{0,1}\left( \upsilon \right)
\varphi \\ 
+\mathrm{i}M_4\beta ^{\left[ 4\right] }U_{0,1}\left( \upsilon \right)
\varphi +\widehat{M}U_{0,1}\left( \upsilon \right) \varphi
\end{array}
\right) \mbox{.} 
\]

Therefore:

\[
\left( 
\begin{array}{c}
U_{0,1}^{\dagger }\left( \upsilon \right) \frac 1{\mathrm{c}}\left( \partial
_tU_{0,1}\left( \upsilon \right) \right) +U_{0,1}^{\dagger }\left( \upsilon
\right) U_{0,1}\left( \upsilon \right) \frac 1{\mathrm{c}}\partial _t \\ 
+\mathrm{i}\Theta _0U_{0,1}^{\dagger }\left( \upsilon \right) U_{0,1}\left(
\upsilon \right) +\mathrm{i}\Upsilon _0U_{0,1}^{\dagger }\left( \upsilon
\right) \gamma ^{\left[ 5\right] }U_{0,1}\left( \upsilon \right)
\end{array}
\right) \varphi = 
\]

\[
=\left( 
\begin{array}{c}
U_{0,1}^{\dagger }\left( \upsilon \right) \beta ^{\left[ 1\right] }\left( 
\begin{array}{c}
\left( \partial _1U_{0,1}\left( \upsilon \right) \right) +U_{0,1}\left(
\upsilon \right) \partial _1 \\ 
+\mathrm{i}\Theta _1U_{0,1}\left( \upsilon \right) +\mathrm{i}\Upsilon
_1\gamma ^{\left[ 5\right] }U_{0,1}\left( \upsilon \right)
\end{array}
\right) + \\ 
+U_{0,1}^{\dagger }\left( \upsilon \right) \beta ^{\left[ 2\right] }\left( 
\begin{array}{c}
\left( \partial _2U_{0,1}\left( \upsilon \right) \right) +U_{0,1}\left(
\upsilon \right) \partial _2 \\ 
+\mathrm{i}\Theta _2U_{0,1}\left( \upsilon \right) +\mathrm{i}\Upsilon
_2\gamma ^{\left[ 5\right] }U_{0,1}\left( \upsilon \right)
\end{array}
\right) + \\ 
+U_{0,1}^{\dagger }\left( \upsilon \right) \beta ^{\left[ 3\right] }\left( 
\begin{array}{c}
\left( \partial _3U_{0,1}\left( \upsilon \right) \right) +U_{0,1}\left(
\upsilon \right) \partial _3 \\ 
+\mathrm{i}\Theta _3U_{0,1}\left( \upsilon \right) +\mathrm{i}\Upsilon
_3\gamma ^{\left[ 5\right] }U_{0,1}\left( \upsilon \right)
\end{array}
\right) + \\ 
+\mathrm{i}M_0U_{0,1}^{\dagger }\left( \upsilon \right) \gamma ^{\left[
0\right] }U_{0,1}\left( \upsilon \right) \\ 
+\mathrm{i}M_4U_{0,1}^{\dagger }\left( \upsilon \right) \beta ^{\left[
4\right] }U_{0,1}\left( \upsilon \right) +U_{0,1}^{\dagger }\left( \upsilon
\right) \widehat{M}U_{0,1}\left( \upsilon \right)
\end{array}
\right) \varphi \mbox{.} 
\]

Since (about a sum of matrix and number in \cite{PT!0}):

\begin{eqnarray*}
U_{0,1}^{\dagger }\left( \upsilon \right) U_{0,1}\left( \upsilon \right)
&=&\left( \cosh 2\upsilon -\beta ^{\left[ 1\right] }\sinh 2\upsilon \right) %
\mbox{,} \\
U_{0,1}^{\dagger }\left( \upsilon \right) &=&\left( \cosh 2\upsilon +\beta
^{\left[ 1\right] }\sinh 2\upsilon \right) U_{0,1}^{-1}\left( \upsilon
\right) \mbox{,} \\
U_{0,1}^{\dagger }\left( \upsilon \right) \beta ^{\left[ 1\right] }
&=&\left( \beta ^{\left[ 1\right] }\cosh 2\upsilon -\sinh 2\upsilon \right)
U_{0,1}^{-1}\left( \upsilon \right) \mbox{,} \\
U_{0,1}^{\dagger }\left( \upsilon \right) \beta ^{\left[ 2\right] } &=&\beta
^{\left[ 2\right] }U_{0,1}^{-1}\left( \upsilon \right) \mbox{,} \\
U_{0,1}^{\dagger }\left( \upsilon \right) \beta ^{\left[ 3\right] } &=&\beta
^{\left[ 3\right] }U_{0,1}^{-1}\left( \upsilon \right) \mbox{,} \\
U_{0,1}^{\dagger }\left( \upsilon \right) \gamma ^{\left[ 0\right]
}U_{0,1}\left( \upsilon \right) &=&\gamma ^{\left[ 0\right] }\mbox{,} \\
U_{0,1}^{\dagger }\left( \upsilon \right) \beta ^{\left[ 4\right]
}U_{0,1}\left( \upsilon \right) &=&\beta ^{\left[ 4\right] }
\end{eqnarray*}

then

\[
\left( 
\begin{array}{c}
\left( \cosh 2\upsilon +\beta ^{\left[ 1\right] }\sinh 2\upsilon \right)
U_{0,1}^{-1}\left( \upsilon \right) \frac 1{\mathrm{c}}\left( \partial
_tU_{0,1}\left( \upsilon \right) \right) \\ 
+\left( \cosh 2\upsilon -\beta ^{\left[ 1\right] }\sinh 2\upsilon \right)
\frac 1{\mathrm{c}}\partial _t \\ 
+\mathrm{i}\Theta _0\left( \cosh 2\upsilon -\beta ^{\left[ 1\right] }\sinh
2\upsilon \right) \\ 
+\mathrm{i}\Upsilon _0U_{0,1}^{\dagger }\left( \upsilon \right) \gamma
^{\left[ 5\right] }U_{0,1}\left( \upsilon \right)
\end{array}
\right) \varphi = 
\]

\[
=\left( 
\begin{array}{c}
\left( \beta ^{\left[ 1\right] }\cosh 2\upsilon -\sinh 2\upsilon \right)
\cdot \\ 
\cdot U_{0,1}^{-1}\left( \upsilon \right) \left( 
\begin{array}{c}
\left( \partial _1U_{0,1}\left( \upsilon \right) \right) +U_{0,1}\left(
\upsilon \right) \partial _1 \\ 
+\mathrm{i}\Theta _1U_{0,1}\left( \upsilon \right) +\mathrm{i}\Upsilon
_1\gamma ^{\left[ 5\right] }U_{0,1}\left( \upsilon \right)
\end{array}
\right) + \\ 
+\beta ^{\left[ 2\right] }U_{0,1}^{-1}\left( \upsilon \right) \left( 
\begin{array}{c}
\left( \partial _2U_{0,1}\left( \upsilon \right) \right) +U_{0,1}\left(
\upsilon \right) \partial _2 \\ 
+\mathrm{i}\Theta _2U_{0,1}\left( \upsilon \right) +\mathrm{i}\Upsilon
_2\gamma ^{\left[ 5\right] }U_{0,1}\left( \upsilon \right)
\end{array}
\right) + \\ 
+\beta ^{\left[ 3\right] }U_{0,1}^{-1}\left( \upsilon \right) \left( 
\begin{array}{c}
\left( \partial _3U_{0,1}\left( \upsilon \right) \right) +U_{0,1}\left(
\upsilon \right) \partial _3 \\ 
+\mathrm{i}\Theta _3U_{0,1}\left( \upsilon \right) +\mathrm{i}\Upsilon
_3\gamma ^{\left[ 5\right] }U_{0,1}\left( \upsilon \right)
\end{array}
\right) + \\ 
+\mathrm{i}M_0\gamma ^{\left[ 0\right] }+\mathrm{i}M_4\beta ^{\left[
4\right] }+\widehat{M}^{\prime \prime }
\end{array}
\right) \varphi \mbox{.} 
\]

Hence:

\[
\left( 
\begin{array}{c}
\left( \cosh 2\upsilon +\beta ^{\left[ 1\right] }\sinh 2\upsilon \right)
U_{0,1}^{-1}\left( \upsilon \right) \frac 1{\mathrm{c}}\left( \partial
_tU_{0,1}\left( \upsilon \right) \right) \\ 
+\left( \cosh 2\upsilon -\beta ^{\left[ 1\right] }\sinh 2\upsilon \right)
\frac 1{\mathrm{c}}\partial _t \\ 
+\mathrm{i}\Theta _0\left( \cosh 2\upsilon -\beta ^{\left[ 1\right] }\sinh
2\upsilon \right) \\ 
+\mathrm{i}\Upsilon _0U_{0,1}^{\dagger }\left( \upsilon \right) \gamma
^{\left[ 5\right] }U_{0,1}\left( \upsilon \right)
\end{array}
\right) \varphi = 
\]

\[
=\left( 
\begin{array}{c}
\left( \beta ^{\left[ 1\right] }\cosh 2\upsilon -\sinh 2\upsilon \right)
\cdot \\ 
\cdot \left( 
\begin{array}{c}
U_{0,1}^{-1}\left( \upsilon \right) \left( \partial _1U_{0,1}\left( \upsilon
\right) \right) +U_{0,1}^{-1}\left( \upsilon \right) U_{0,1}\left( \upsilon
\right) \partial _1 \\ 
+\mathrm{i}\Theta _1U_{0,1}^{-1}\left( \upsilon \right) U_{0,1}\left(
\upsilon \right) +\mathrm{i}\Upsilon _1U_{0,1}^{-1}\left( \upsilon \right)
\gamma ^{\left[ 5\right] }U_{0,1}\left( \upsilon \right)
\end{array}
\right) + \\ 
+\beta ^{\left[ 2\right] }\left( 
\begin{array}{c}
U_{0,1}^{-1}\left( \upsilon \right) \left( \partial _2U_{0,1}\left( \upsilon
\right) \right) +U_{0,1}^{-1}\left( \upsilon \right) U_{0,1}\left( \upsilon
\right) \partial _2 \\ 
+\mathrm{i}\Theta _2U_{0,1}^{-1}\left( \upsilon \right) U_{0,1}\left(
\upsilon \right) +\mathrm{i}\Upsilon _2U_{0,1}^{-1}\left( \upsilon \right)
\gamma ^{\left[ 5\right] }U_{0,1}\left( \upsilon \right)
\end{array}
\right) + \\ 
+\beta ^{\left[ 3\right] }\left( 
\begin{array}{c}
U_{0,1}^{-1}\left( \upsilon \right) \left( \partial _3U_{0,1}\left( \upsilon
\right) \right) +U_{0,1}^{-1}\left( \upsilon \right) U_{0,1}\left( \upsilon
\right) \partial _3 \\ 
+\mathrm{i}\Theta _3U_{0,1}^{-1}\left( \upsilon \right) U_{0,1}\left(
\upsilon \right) +\mathrm{i}\Upsilon _3U_{0,1}^{-1}\left( \upsilon \right)
\gamma ^{\left[ 5\right] }U_{0,1}\left( \upsilon \right)
\end{array}
\right) + \\ 
+\mathrm{i}M_0\gamma ^{\left[ 0\right] }+\mathrm{i}M_4\beta ^{\left[
4\right] }+\widehat{M}^{\prime \prime }
\end{array}
\right) \varphi \mbox{.} 
\]

Because

\begin{eqnarray*}
U_{0,1}^{-1}\left( \upsilon \right) U_{0,1}\left( \upsilon \right) &=&1_4%
\mbox{,} \\
U_{0,1}^{-1}\left( \upsilon \right) \gamma ^{\left[ 5\right] }U_{0,1}\left(
\upsilon \right) &=&\gamma ^{\left[ 5\right] }\mbox{,} \\
U_{0,1}^{\dagger }\left( \upsilon \right) \gamma ^{\left[ 5\right]
}U_{0,1}\left( \upsilon \right) &=&\gamma ^{\left[ 5\right] }\left( \cosh
2\upsilon -\beta ^{\left[ 1\right] }\sinh 2\upsilon \right)
\end{eqnarray*}

then

\[
\left( 
\begin{array}{c}
\left( \cosh 2\upsilon +\beta ^{\left[ 1\right] }\sinh 2\upsilon \right)
U_{0,1}^{-1}\left( \upsilon \right) \frac 1{\mathrm{c}}\partial
_tU_{0,1}\left( \upsilon \right) \\ 
+\left( \cosh 2\upsilon -\beta ^{\left[ 1\right] }\sinh 2\upsilon \right)
\frac 1{\mathrm{c}}\partial _t \\ 
+\mathrm{i}\Theta _0\left( \cosh 2\upsilon -\beta ^{\left[ 1\right] }\sinh
2\upsilon \right) \\ 
+\mathrm{i}\Upsilon _0\gamma ^{\left[ 5\right] }\left( \cosh 2\upsilon
-\beta ^{\left[ 1\right] }\sinh 2\upsilon \right)
\end{array}
\right) \varphi = 
\]

\[
=\left( 
\begin{array}{c}
\left( \beta ^{\left[ 1\right] }\cosh 2\upsilon -\sinh 2\upsilon \right)
\cdot \\ 
\cdot \left( \partial _1+U_{0,1}^{-1}\left( \upsilon \right) \left( \partial
_1U_{0,1}\left( \upsilon \right) \right) +\mathrm{i}\Theta _1+\mathrm{i}%
\Upsilon _1\gamma ^{\left[ 5\right] }\right) + \\ 
+\beta ^{\left[ 2\right] }\left( \partial _2+U_{0,1}^{-1}\left( \upsilon
\right) \left( \partial _2U_{0,1}\left( \upsilon \right) \right) +\mathrm{i}%
\Theta _2+\mathrm{i}\Upsilon _2\gamma ^{\left[ 5\right] }\right) + \\ 
+\beta ^{\left[ 3\right] }\left( \partial _3+U_{0,1}^{-1}\left( \upsilon
\right) \left( \partial _3U_{0,1}\left( \upsilon \right) \right) +\mathrm{i}%
\Theta _3+\mathrm{i}\Upsilon _3\gamma ^{\left[ 5\right] }\right) + \\ 
+\mathrm{i}M_0\gamma ^{\left[ 0\right] }+\mathrm{i}M_4\beta ^{\left[
4\right] }+\widehat{M}^{\prime \prime }
\end{array}
\right) \varphi \mbox{.} 
\]

Hence:

\[
\left( 
\begin{array}{c}
\cosh 2\upsilon \cdot U_{0,1}^{-1}\left( \upsilon \right) \frac 1{\mathrm{c}%
}\partial _tU_{0,1}\left( \upsilon \right) +\cosh 2\upsilon \cdot \frac 1{%
\mathrm{c}}\partial _t+\sinh 2\upsilon \cdot \partial _1 \\ 
+\mathrm{i}\Theta _0\cosh 2\upsilon +\mathrm{i}\Upsilon _0\gamma ^{\left[
5\right] }\cosh 2\upsilon \\ 
+\sinh 2\upsilon \cdot U_{0,1}^{-1}\left( \upsilon \right) \left( \partial
_1U_{0,1}\left( \upsilon \right) \right) \\ 
+\mathrm{i}\Theta _1\sinh 2\upsilon \\ 
+\mathrm{i}\sinh 2\upsilon \cdot \Upsilon _1\gamma ^{\left[ 5\right] } \\ 
+\beta ^{\left[ 1\right] }\sinh 2\upsilon \cdot U_{0,1}^{-1}\left( \upsilon
\right) \frac 1{\mathrm{c}}\partial _tU_{0,1}\left( \upsilon \right) -\beta
^{\left[ 1\right] }\sinh 2\upsilon \cdot \frac 1{\mathrm{c}}\partial _t \\ 
-\mathrm{i}\beta ^{\left[ 1\right] }\Theta _0\sinh 2\upsilon -\mathrm{i}%
\beta ^{\left[ 1\right] }\Upsilon _0\gamma ^{\left[ 5\right] }\sinh
2\upsilon -\beta ^{\left[ 1\right] }\cosh 2\upsilon \cdot \partial _1 \\ 
-\beta ^{\left[ 1\right] }\cosh 2\upsilon \cdot U_{0,1}^{-1}\left( \upsilon
\right) \left( \partial _1U_{0,1}\left( \upsilon \right) \right) \\ 
-\mathrm{i}\beta ^{\left[ 1\right] }\Theta _1\cosh 2\upsilon \\ 
-\mathrm{i}\beta ^{\left[ 1\right] }\cosh 2\upsilon \cdot \Upsilon _1\gamma
^{\left[ 5\right] } \\ 
-\beta ^{\left[ 2\right] }\left( \partial _2+U_{0,1}^{-1}\left( \upsilon
\right) \left( \partial _2U_{0,1}\left( \upsilon \right) \right) +\mathrm{i}%
\Theta _2+\mathrm{i}\Upsilon _2\gamma ^{\left[ 5\right] }\right) \\ 
-\beta ^{\left[ 3\right] }\left( \partial _3+U_{0,1}^{-1}\left( \upsilon
\right) \left( \partial _3U_{0,1}\left( \upsilon \right) \right) +\mathrm{i}%
\Theta _3+\mathrm{i}\Upsilon _3\gamma ^{\left[ 5\right] }\right) \\ 
-\mathrm{i}M_0\gamma ^{\left[ 0\right] }-\mathrm{i}M_4\beta ^{\left[
4\right] }-\widehat{M}^{\prime \prime }
\end{array}
\right) \varphi =0\mbox{.} 
\]

Therefore,

\begin{equation}
\left( 
\begin{array}{c}
U_{0,1}^{-1}\left( \upsilon \right) \left( \cosh 2\upsilon \cdot \frac 1{%
\mathrm{c}}\partial _t+\sinh 2\upsilon \cdot \partial _1\right)
U_{0,1}\left( \upsilon \right) \\ 
+\left( \cosh 2\upsilon \cdot \frac 1{\mathrm{c}}\partial _t+\sinh 2\upsilon
\cdot \partial _1\right) \\ 
+\mathrm{i}\left( \Theta _0\cosh 2\upsilon +\Theta _1\sinh 2\upsilon \right)
\\ 
+\mathrm{i}\left( \Upsilon _0\cosh 2\upsilon +\sinh 2\upsilon \cdot \Upsilon
_1\right) \gamma ^{\left[ 5\right] } \\ 
-\beta ^{\left[ 1\right] }\left( 
\begin{array}{c}
U_{0,1}^{-1}\left( \upsilon \right) \left( \cosh 2\upsilon \cdot \partial
_1+\sinh 2\upsilon \cdot \frac 1{\mathrm{c}}\partial _t\right) U_{0,1}\left(
\upsilon \right) \\ 
+\left( \cosh 2\upsilon \cdot \partial _1+\sinh 2\upsilon \cdot \frac 1{%
\mathrm{c}}\partial _t\right) \\ 
+\mathrm{i}\left( \Theta _1\cosh 2\upsilon +\Theta _0\sinh 2\upsilon \right)
\\ 
+\mathrm{i}\left( \Upsilon _1\cosh 2\upsilon +\Upsilon _0\sinh 2\upsilon
\right) \gamma ^{\left[ 5\right] }
\end{array}
\right) \\ 
-\beta ^{\left[ 2\right] }\left( \partial _2+U_{0,1}^{-1}\left( \upsilon
\right) \left( \partial _2U_{0,1}\left( \upsilon \right) \right) +\mathrm{i}%
\Theta _2+\mathrm{i}\Upsilon _2\gamma ^{\left[ 5\right] }\right) \\ 
-\beta ^{\left[ 3\right] }\left( \partial _3+U_{0,1}^{-1}\left( \upsilon
\right) \left( \partial _3U_{0,1}\left( \upsilon \right) \right) +\mathrm{i}%
\Theta _3+\mathrm{i}\Upsilon _3\gamma ^{\left[ 5\right] }\right) \\ 
-\mathrm{i}M_0\gamma ^{\left[ 0\right] }-\mathrm{i}M_4\beta ^{\left[
4\right] }-\widehat{M}^{\prime \prime }
\end{array}
\right) \varphi =0  \label{ham6}
\end{equation}

Let $t^{\prime }$ and $x_1^{\prime }$ are elements of other coordinates
system such that:

\begin{eqnarray*}
\ &&\frac{\partial x_1}{\partial x_1^{\prime }}=\cosh 2\upsilon \mbox{,} \\
\ &&\frac{\partial t}{\partial x_1^{\prime }}=\frac 1{\mathrm{c}}\sinh
2\upsilon \mbox{,} \\
\ &&\frac{\partial x_1}{\partial t^{\prime }}=\mathrm{c}\sinh 2\upsilon %
\mbox{.} \\
\ &&\frac{\partial t}{\partial t^{\prime }}=\cosh 2\upsilon \mbox{,} \\
\ &&\frac{\partial x_2}{\partial t^{\prime }}=\frac{\partial x_3}{\partial
t^{\prime }}=\frac{\partial x_2}{\partial x_1^{\prime }}=\frac{\partial x_3}{%
\partial x_1^{\prime }}=0\mbox{.}
\end{eqnarray*}

Hence:

\begin{eqnarray*}
\partial _t^{\prime }\stackrel{def}{=}\frac \partial {\partial t^{\prime }}
&=&\frac \partial {\partial t}\frac{\partial t}{\partial t^{\prime }}+\frac
\partial {\partial x_1}\frac{\partial x_1}{\partial t^{\prime }}+\frac
\partial {\partial x_2}\frac{\partial x_2}{\partial t^{\prime }}+\frac
\partial {\partial x_3}\frac{\partial x_3}{\partial i^{\prime }} \\
\ &=&\cosh 2\upsilon \cdot \frac \partial {\partial t}+\mathrm{c}\sinh
2\upsilon \cdot \frac \partial {\partial x_1} \\
\ &=&\cosh 2\upsilon \cdot \partial _t+\mathrm{c}\sinh 2\upsilon \cdot
\partial _1\mbox{,}
\end{eqnarray*}

That is:

\[
\frac 1{\mathrm{c}}\partial _t^{\prime }=\frac 1{\mathrm{c}}\cosh 2\upsilon
\cdot \partial _t+\sinh 2\upsilon \cdot \partial _1\mbox{.} 
\]

And

\begin{eqnarray*}
\partial _1^{\prime }\stackrel{def}{=}\frac \partial {\partial x_1^{\prime
}} &=&\frac \partial {\partial t}\frac{\partial t}{\partial x_1^{\prime }}%
+\frac \partial {\partial x_1}\frac{\partial x_1}{\partial x_1^{\prime }}%
+\frac \partial {\partial x_2}\frac{\partial x_2}{\partial x_1^{\prime }}%
+\frac \partial {\partial x_3}\frac{\partial x_3}{\partial x_1^{\prime }} \\
\ &=&\cosh 2\upsilon \cdot \frac \partial {\partial x_1}+\sinh 2\upsilon
\cdot \frac 1{\mathrm{c}}\frac \partial {\partial t} \\
\ &=&\cosh 2\upsilon \cdot \partial _1+\sinh 2\upsilon \cdot \frac 1{\mathrm{%
c}}\partial _t.
\end{eqnarray*}

Therefore from (\ref{ham6}):

\[
\left( 
\begin{array}{c}
+\frac 1{\mathrm{c}}\partial _t^{\prime }+U_{0,1}^{-1}\left( \upsilon
\right) \frac 1{\mathrm{c}}\partial _t^{\prime }U_{0,1}\left( \upsilon
\right) +\mathrm{i}\Theta _0^{\prime \prime }+\mathrm{i}\Upsilon _0^{\prime
\prime }\gamma ^{\left[ 5\right] } \\ 
-\beta ^{\left[ 1\right] }\left( +\partial _1^{\prime }+U_{0,1}^{-1}\left(
\upsilon \right) \partial _1^{\prime }U_{0,1}\left( \upsilon \right) +%
\mathrm{i}\Theta _1^{\prime \prime }+\mathrm{i}\Upsilon _1^{\prime \prime
}\gamma ^{\left[ 5\right] }\right) \\ 
-\beta ^{\left[ 2\right] }\left( \partial _2+U_{0,1}^{-1}\left( \upsilon
\right) \left( \partial _2U_{0,1}\left( \upsilon \right) \right) +\mathrm{i}%
\Theta _2+\mathrm{i}\Upsilon _2\gamma ^{\left[ 5\right] }\right) \\ 
-\beta ^{\left[ 3\right] }\left( \partial _3+U_{0,1}^{-1}\left( \upsilon
\right) \left( \partial _3U_{0,1}\left( \upsilon \right) \right) +\mathrm{i}%
\Theta _3+\mathrm{i}\Upsilon _3\gamma ^{\left[ 5\right] }\right) \\ 
-\mathrm{i}M_0\gamma ^{\left[ 0\right] }-\mathrm{i}M_4\beta ^{\left[
4\right] }-\widehat{M}^{\prime \prime }
\end{array}
\right) \varphi =0 
\]

with

\[
\begin{array}{c}
\Theta _0^{\prime \prime }\stackrel{def}{=}\Theta _0\cosh 2\upsilon +\Theta
_1\sinh 2\upsilon \mbox{,} \\ 
\Theta _1^{\prime \prime }\stackrel{def}{=}\Theta _1\cosh 2\upsilon +\Theta
_0\sinh 2\upsilon \mbox{,} \\ 
\Upsilon _0^{\prime \prime }\stackrel{def}{=}\Upsilon _0\cosh 2\upsilon
+\sinh 2\upsilon \cdot \Upsilon _1\mbox{,} \\ 
\Upsilon _1^{\prime \prime }\stackrel{def}{=}\Upsilon _1\cosh 2\upsilon
+\Upsilon _0\sinh 2\upsilon \mbox{.}
\end{array}
\]

Therefore, the oscillation between blue and green quarks colors with the
oscillation between up and down quarks curves the space of events in the $t$%
, $x_1$ directions.

Similarly that: matrix

\[
U_{0,2}\left( \phi \right) =\left[ 
\begin{array}{cccc}
\cosh \phi & i\sinh \phi & 0 & 0 \\ 
-i\sinh \phi & \cosh \phi & 0 & 0 \\ 
0 & 0 & \cosh \phi & -i\sinh \phi \\ 
0 & 0 & i\sinh \phi & \cosh \phi
\end{array}
\right] 
\]

with an arbitrary real function $\phi \left( t,x_1,x_2,x_3\right) $
describes the oscillation between blue and red quarks colors with the
oscillation between up and down quarks which curves the space of events in
the $t$, $x_2$ directions. And matrix

\[
U_{0,3}\left( \iota \right) =\left[ 
\begin{array}{cccc}
\cosh \iota +\sinh \iota & 0 & 0 & 0 \\ 
0 & \cosh \iota -\sinh \iota & 0 & 0 \\ 
0 & 0 & \cosh \iota -\sinh \iota & 0 \\ 
0 & 0 & 0 & \cosh \iota +\sinh \iota
\end{array}
\right] 
\]

with an arbitrary real function $\iota \left( t,x_1,x_2,x_3\right) $
describes the oscillation between green and red quarks colors with the
oscillation between up and down quarks which curves the space of events in
the $t$, $x_3$ directions.

Now, because

$\iota \left( t,x_1,x_2,x_3\right) $, $\phi \left( t,x_1,x_2,x_3\right) $, $%
\upsilon =\upsilon \left( t,x_1,x_2,x_3\right) $, $\varsigma \left(
t,x_1,x_2,x_3\right) $,

$\vartheta \left( t,x_1,x_2,x_3\right) $, $\alpha \left(
t,x_1,x_2,x_3\right) $

are arbitrary real functions then for coordinates $X_\nu $ of the curved
time-space the formula 2, page 155 from \cite{Ein} is fulfils:

\[
dX_\nu =\sum_{\sigma =0}^3a_{\nu \sigma }dx_\sigma 
\]
with real functions $a_{\nu \sigma }\left( t,x_1,x_2,x_3\right) $.

Hence, the square of the linear element $ds^2$ obeys to formula 3, page 155:

\[
ds^2=\sum_{\sigma =0}^3\sum_{\tau =0}^3g_{\sigma \tau }dx_\sigma dx_\tau 
\]

with real functions $g_{\sigma \tau }\left( t,x_1,x_2,x_3\right) $ such that 
$g_{\sigma \tau }=g_{\tau \sigma }$. And then the Einstein deductions from 
\cite{Ein} can be reiterated:

The equation of a geodesic line between points $P$ and $P^{\prime }$ is the
following (formula 20, page 167):

\[
\delta \int_P^{P^{\prime }}ds=0\mbox{.} 
\]

Hence with a parameter $\lambda $:

\[
\delta \int_{\lambda _1}^{\lambda _2}\sqrt{\sum_{\sigma =0}^3\sum_{\tau
=0}^3g_{\sigma \tau }\frac{dx_\sigma }{d\lambda }\frac{dx_\tau }{d\lambda }}%
d\lambda =0\mbox{.} 
\]

If denote:

\[
F(x_\sigma ,x_\tau ,\frac{dx_\sigma }{d\lambda },\frac{dx_\tau }{d\lambda })%
\stackrel{def}{=}\sqrt{\sum_{\sigma =0}^3\sum_{\tau =0}^3g_{\sigma \tau }%
\frac{dx_\sigma }{d\lambda }\frac{dx_\tau }{d\lambda }} 
\]

then in accordance with the Euler-Lagrange variational principle the
geodesic line equation is the following:

\[
\frac{\partial F}{\partial x_\sigma }-\frac d{d\lambda }\frac{\partial F}{%
\partial \left( \frac{dx_\sigma }{d\lambda }\right) }=0\mbox{.} 
\]

Because

\begin{eqnarray*}
\frac{\partial F}{\partial x_\sigma } &=&\frac \partial {\partial x_\sigma
}\left( \sqrt{\sum_{\nu =0}^3\sum_{\mu =0}^3g_{\nu \mu }\frac{dx_\nu }{%
d\lambda }\frac{dx_\mu }{d\lambda }}\right) \\
&=&\frac 12\frac 1{\sqrt{\sum_{\nu =0}^3\sum_{\mu =0}^3g_{\nu \mu }\frac{%
dx_\nu }{d\lambda }\frac{dx_\mu }{d\lambda }}}\sum_{\nu =0}^3\sum_{\mu =0}^3%
\frac{\partial g_{\nu \mu }}{\partial x_\sigma }\frac{dx_\nu }{d\lambda }%
\frac{dx_\mu }{d\lambda }
\end{eqnarray*}

and

\begin{eqnarray*}
\frac d{d\lambda }\frac{\partial F}{\partial \left( \frac{dx_\sigma }{%
d\lambda }\right) } &=&\frac d{d\lambda }\frac \partial {\partial \left( 
\frac{dx_\sigma }{d\lambda }\right) }\left( \sqrt{\sum_{\nu =0}^3\sum_{\mu
=0}^3g_{\nu \mu }\frac{dx_\nu }{d\lambda }\frac{dx_\mu }{d\lambda }}\right)
\\
&=&\frac d{d\lambda }\frac 1{\sqrt{\sum_{\nu =0}^3\sum_{\mu =0}^3g_{\nu \mu }%
\frac{dx_\nu }{d\lambda }\frac{dx_\mu }{d\lambda }}}\sum_{\mu =0}^3g_{\sigma
\mu }\frac{dx_\mu }{d\lambda }\mbox{.}
\end{eqnarray*}

then the equation of geodesic line is the following:

\begin{eqnarray*}
&&\frac 12\frac 1{\sqrt{\sum_{\nu =0}^3\sum_{\mu =0}^3g_{\nu \mu }\frac{%
dx_\nu }{d\lambda }\frac{dx_\mu }{d\lambda }}}\sum_{\nu =0}^3\sum_{\mu =0}^3%
\frac{\partial g_{\nu \mu }}{\partial x_\sigma }\frac{dx_\nu }{d\lambda }%
\frac{dx_\mu }{d\lambda } \\
&&-\frac d{d\lambda }\frac 1{\sqrt{\sum_{\nu =0}^3\sum_{\mu =0}^3g_{\nu \mu }%
\frac{dx_\nu }{d\lambda }\frac{dx_\mu }{d\lambda }}}\sum_{\mu =0}^3g_{\sigma
\mu }\frac{dx_\mu }{d\lambda }=0\mbox{.}
\end{eqnarray*}

if $\lambda =s$ then

\[
\sum_{\nu =0}^3\sum_{\mu =0}^3g_{\nu \mu }\frac{dx_\nu }{d\lambda }\frac{%
dx_\mu }{d\lambda }=\frac{\sum_{\nu =0}^3\sum_{\mu =0}^3g_{\nu \mu }dx_\nu
dx_\mu }{d\lambda ^2}=\frac{ds^2}{ds^2}=1\mbox{,} 
\]

Hence

\begin{eqnarray*}
&&\frac d{d\lambda }\frac 1{\sqrt{\sum_{\nu =0}^3\sum_{\mu =0}^3g_{\nu \mu }%
\frac{dx_\nu }{d\lambda }\frac{dx_\mu }{d\lambda }}}\sum_{\mu =0}^3g_{\sigma
\mu }\frac{dx_\mu }{d\lambda }=\frac d{d\lambda }\sum_{\mu =0}^3g_{\sigma
\mu }\frac{dx_\mu }{d\lambda } \\
&=&\sum_{\mu =0}^3\left( \left( \frac d{d\lambda }g_{\sigma \mu }\right) 
\frac{dx_\mu }{d\lambda }+g_{\sigma \mu }\frac d{d\lambda }\frac{dx_\mu }{%
d\lambda }\right) \\
&=&\sum_{\mu =0}^3\left( \sum_{\tau =0}^3\frac{\partial g_{\sigma \mu }}{%
\partial x_\tau }\frac{\partial x_\tau }{d\lambda }\frac{dx_\mu }{d\lambda }%
+g_{\sigma \mu }\frac d{d\lambda }\frac{dx_\mu }{d\lambda }\right) \\
&=&\sum_{\mu =0}^3\left( \sum_{\tau =0}^3\frac{\partial g_{\sigma \mu }}{%
\partial x_\tau }\frac{\partial x_\tau }{ds}\frac{dx_\mu }{ds}+g_{\sigma \mu
}\frac d{ds}\frac{dx_\mu }{ds}\right) \mbox{.}
\end{eqnarray*}

Therefore, the geodesic line equation is the following:

\[
\sum_{\mu =0}^3\left( g_{\sigma \mu }\frac{d^2x_\mu }{ds^2}+\sum_{\tau =1}^3%
\frac{\partial g_{\sigma \mu }}{\partial x_\tau }\frac{\partial x_\tau }{ds}%
\frac{dx_\mu }{ds}\right) -\frac 12\sum_{\mu =0}^3\sum_{\nu =0}^3\frac{%
\partial g_{\nu \mu }}{\partial x_\sigma }\frac{dx_\nu }{ds}\frac{dx_\mu }{ds%
}=0\mbox{.} 
\]

$=============$

\end{document}